# Bringing Robustness to End-User Programming

Mickaël Baron, Patrick Girard
*Laboratoire d'Informatique Scientifique et Industrielle, ENSMA,*
*1 rue Clément Ader, 86961 Futuroscope Chasseneuil*
*http://www.lisi.ensma.fr/ihm*
*{baron, girard}@ensma.fr*

## Abstract

*In some cases, end-user programming allows the design of stand-alone applications. But none of the existing approaches is concerned by safety aspects of programming. Heavy techniques exist to develop safe applications, particularly in non-interactive domains. They involve software engineering techniques, and sometimes, formal methods. All these techniques are very far from end-users. Our idea is to let this part to experts, and to connect end-user programming onto this safe conventional development. Starting from an existing functional core, we built an interactive end-user programming environment called* ***GenBuild,*** *which allows designing interactive stand-alone applications.* ***GenBuild*** *allows the verification of some properties that are a first step towards the development of safe end-user programming.*

## 1. Introduction

Development tools used for end-users have expanded rapidly. A lot of research tools, but also some commercial products, allow end-users to develop stand-alone applications. The main characteristic of "end-users" in this context is that they do not have programming knowledge. They program occasionally, and this is not the most important part of their activity. They may be experienced computer users, but they are not experienced in conventional programming languages, such as C, Ada, or Java. They may use end-user programming to assume some development tasks, such as *Visual programming* and *Programming by example.* They authorize a fast design of application. Moreover, they permit automating repeated tasks[1].

In fact all these techniques reduce the complexity of interactive application programming, but they do not necessarily ensure the correctness of the resulting applications. Most works on end-user programming focus on expressive power of such techniques, and on intuitivity and simplicity of use for end-users. None of them, at our knowledge, is concerned with safety aspects of end-user development. Is the resulting program correct? Can we ensure that it gives a right solution? Does it crash? Is it complete?

Answering these questions is very difficult in a general way. Nevertheless, we think that researches on end-user programming must be concerned by giving help to end-users in order to program correct applications.

End-user programming deals with highly interactive programs. One of the major assumptions of the end-user programming field is that users can do many things with their applications, and end-user programming can be seen as an optimization of interactive application usage. In the Human-Computer Interaction (HCI) field, separating the interface from the functional core is considered as the basis of a good design. If we split the application development into two stages, one for the functional core, and the other for the interactive part, we can consider the second one in the perspective of end-user programming.

Our approach is the following. Heavy techniques exist to develop safe applications, particularly in non-interactive domains. They involve software engineering techniques, and sometimes, formal methods. All these techniques are very far from end-users. Our idea is to let this part to experts, and to connect end-user programming onto this safe conventional development. Starting from an existing functional core, we built an interactive end-user programming environment that allow the design of interactive stand alone applications.

The purpose of our contribution is to show that end-user programming is not incompatible with safety requirements.

The summary of this paper is the following. In section 2, we define the properties that concern safety in interactive applications. Section 3 deals with the **GenBuild** tool we developed. We present the first module, created by the domain expert, which is connected to a second module allowing an end-user to build interactively a complete application. Section 4 permits to compare our work to some products from research laboratories or commercial products. Last, we give some perspectives of our work.

## 2. Safety for interactive applications

What is software safety? In critical systems, safety has been defined as a collection of properties only related to software design. In interactive systems safety may be defined differently [2]. Because of the specificity of interactive system architectures, where functional core is generally separated from the user interface, we can distinguish two aspects in safety properties: functional core safety and user interface safety. The security properties that concern the functional core allow ensuring that the functional core primitives execute the required result and that there is no execution conflict.

Among the security properties concerning the interface, we can find some that are linked to the system visualization, such as observability, insistence and honesty [3].

- **Observability**: The system makes all pertinent information potentially available to the user;
- **Insistence**: The dialog structure ensures that necessary information is perceived;
- **Honesty**: The dialog structure ensures that users correctly interpret perceived information;

Many other properties may be elected for interactive systems. However, this set of properties seems consistent in our context, and we will restrict our analysis to them.

Following our main idea of separating application programming between an expert domain and an end-user domain, we will show in the next section how different methods can be used to ensure these properties.

## 3. The GenBuild tool

We realized an application called **GenBuild** which means "Generator – Builder". Its main goal is to allow end-users to build interactive application from existing functional cores, with respect of safety. We opted for separating our application into two modules, the first one is called "Generator" and the second one is called "Builder". We made that choice to distinguish the safety aspects of the application concerning the functional core from the safety aspects concerning the interface of the application. Furthermore, the separation permits to distinguish the actors of these tools who have different knowledge. The first actor is a domain expert who understands conventional programming. He/she knows how to implement all the aspects of software design. The second actor is the end-user who has a little or no prior experience in programming. This is a final user who normally uses computer applications, and whose habits are personal works. For example, he/she wishes a unique interface for all the software he/she uses, e.g. the same behavior and the same philosophy. This may lead to a definition of pleasure of working: "Not to learn again a new concept of using for each working tool".

The domain expert uses the "Generator" tool. He/She may write a functional core with a high level of safety. Then an entirely automatic stage allows the Generator tool to extract information from the functional core in order to generate a file. This stage interfaces the functional core with an interactive building environment.

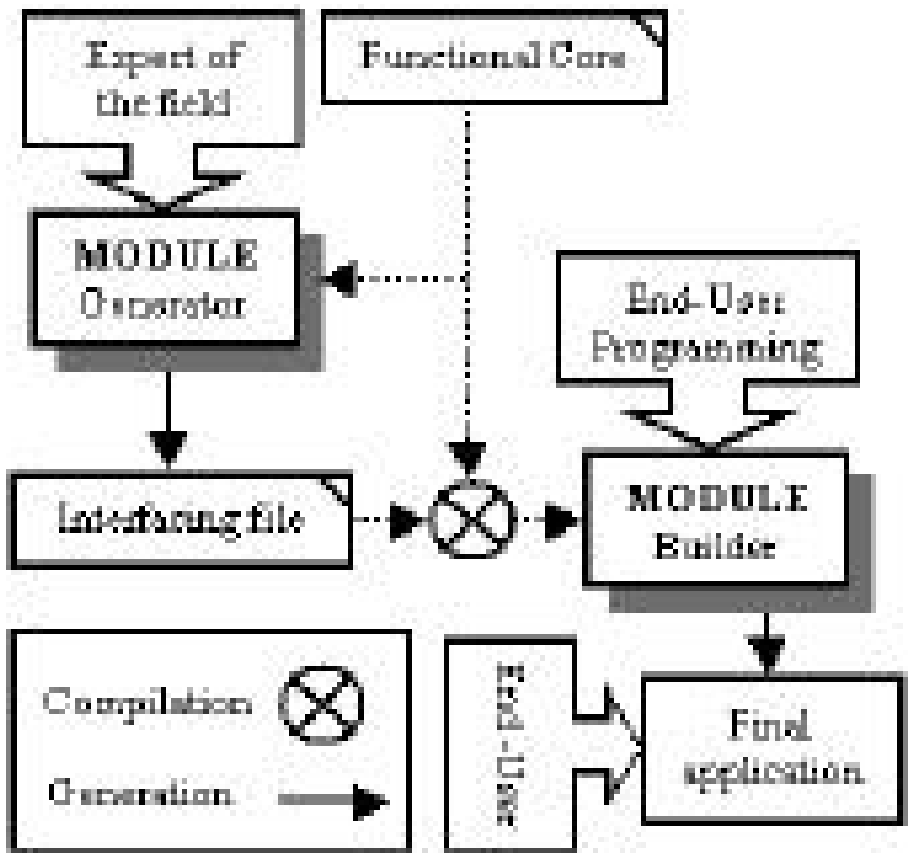

**Figure 1. GenBuild architecture**

The second module is called "Builder". It results from the compilation of the functional core with the interfacing file that comes from the generator module. It is in fact an interface editor, which allows the realization of the final interactive application. The builder user is able to test the functional core by using a default presentation, and more, may alternate the stages of design and test of the final interactive application.

The Figure 1 is a representation of the **GenBuild** architecture. The places for each actor, either the expert programmer or the end-user, are clearly defined

### 3.1. Generator

This module generates standard interactive interfaces from functional cores. So doing, it builds a comprehensive interactive representation for end-users. Firstly, we explain the reason why we chose to let a domain expert develop the functional core and why we want it to write this part in a classical language. To be more concrete, we then detail our example of target application: the **TicTacToe** game. In the second part, we deal with the generation itself.

**The domain expert**: the first step consists in designing a functional core in some classical language. For example this language may be $C^{++}$, Java or Ada, that are all classical languages. We made this choice for functional core robustness reasons. Classical languages have a more important expressive power than any visual language, then it can easily describe the required model. It is important to ensure that the functional core may be used as the basis of interactive development. So, in a first approach, we defined a semi-formal way to modelize the interface of the functional core [4]. So doing, we can exploit this semi-formal description to generate a standard representation, and to have a good basis for end-user development.

**The functional core**: in order to illustrate this explanation, let us describe a concrete example on which we can rely. We chose a well-known simple game, the TicTacToe. This game involves two competitors who play in turns. Winning consists in aligning three symbols on a grid. We restrict our domain to an interactive game between two human players. Figure 2 shows a potential interface for the final application.

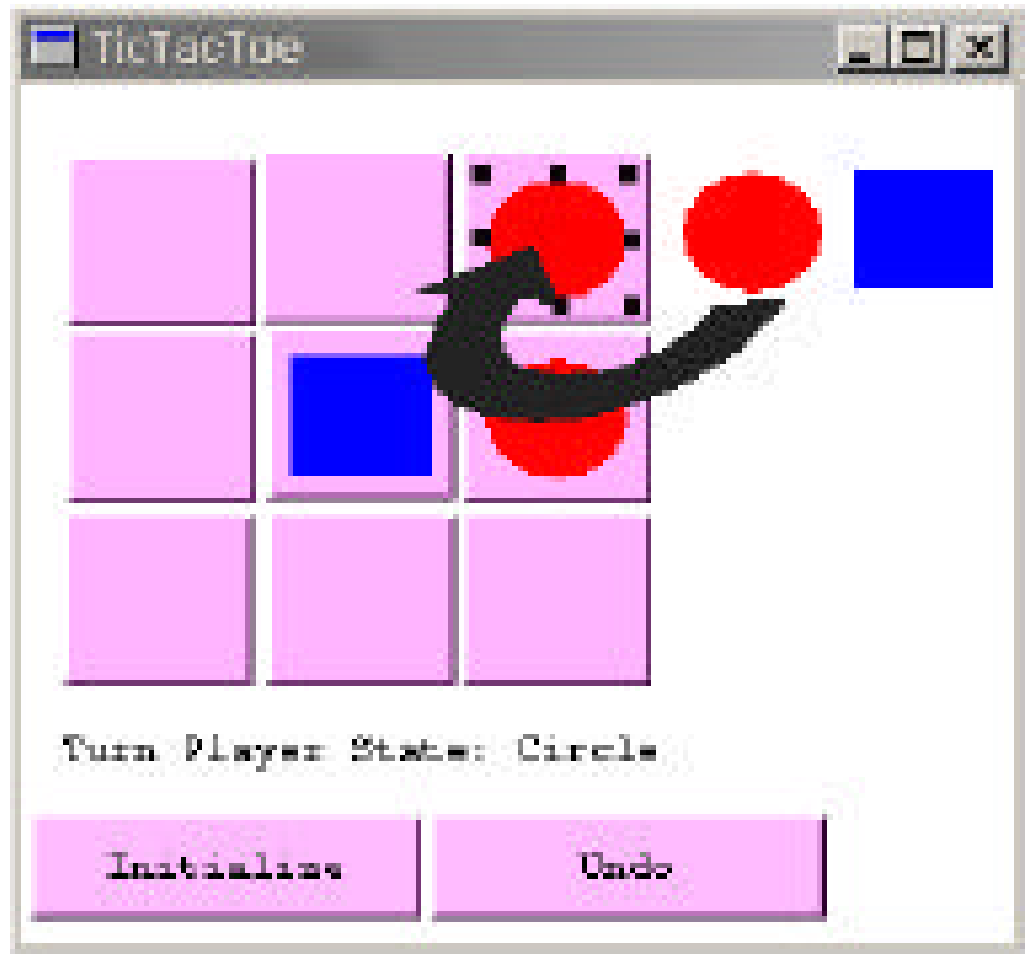

**Figure 2. Interface of TicTacToe game**

The game area is composed of a 3x3 grid, each place contains a symbol corresponding to some player (a plain square or a dot). Users can play by two means. One of the methods consists in clicking directly on the square where the player wants to put down one piece. The other one consists in moving directly the piece onto the grid, using in fact direct manipulation. The application state (Who is playing? Is there a winner?) is represented by textual messages. Two buttons may modify the state of the functional core. A first button puts the application back to its initial state. The second button consists in allowing the user to cancel the last operation he/she made (the UNDO function). In our case it is a matter of canceling the action of selection of one grid square.

Let us concentrate on the functional core. As any functional core, it is a good policy to develop it in an independent way from any HCI perspective. That is to say we must only find into the functions of the application. We chose a classical language, $C^{++}$, which is a popular object-oriented language. We created a class that defines completely the game. We did not define any subclass (squares for example), in order to keep the model very simple, even if such a decomposition would have been better for object-oriented criteria. Beyond the class constructor and an *Initialize* function, we defined the following services, in respect to Fekete's rules [5]:

- **The automatic administration** of the players. Players are represented by two integer values (one or two). At initialization, the game still allows gamer one to play, then it alternates between the two until the end of the game. The number of turns is also administrated by the functional core. The functions are *token_state()* and *turn_count()*;
- **The game itself**. It consists in playing in some given square. The function *play(int i, int j)* permits to modify the functional core according to the active player; the result of this action is that the square (identified by i and j) contains an active player piece.
- **Several functions** return the state of the functional core. They are *win_state()*, *end_state()*, *turn_player_state()* or *token_type(i,j)*; the last function result indicates the content of the i,j grid square.
- **Canceling**. We implemented a systematic canceling of actions on the functional core, with no limitation of the number of undos.

The whole functional core is described with semi-formal specifications. The model uses pre- and post-conditions. For example, Figure 3 gives the specification of the *token_type* function, which returns the state of the square defined by i and j.

```
1   int token_type(int i, int j)
2   // requires: i >= 1 & i <= 3 & j >= 1 & j <= 3
3   // initial value : 0
4   // out value : 0, 1, 2
```

**Figure 3. Semi-formal Specification**

On line two, it shows the pre-condition on in-parameters, which says that i and j must belong to the right interval, in order to ensure not using values out of field. Notice that the syntax must conform the $C^{++}$ language syntax for boolean expressions. It is used just as it is written by the Builder tool. Then, the method initialization is indicated (line 3). Finally, to check the good result of the method, we control the out value (line 4). 0 stands for no piece in the square, while 1 and 2 stand for a piece of the first or the second player, respectively. By applying this formal description to every method of the functional core class, we have elements for verifying the correctness of the interface. The functional core is assumed to respect these conditions on out values.

**Interfacing file**: the "Generator" module generates an interfacing file which represents a standard interactive interface of the functional core. We have seen that it was composed of a class and semi-formal specifications. The module interprets this information in an interface with which the end-user may manipulate the core function. This interface generation is completely automatic: to extract data from the functional core, the generator uses Lex and Yacc tools. Analysing $C^{++}$ and semi-formal specifications from the ".h" file, the generator extracts information written by the domain expert. He/She starts by extracting information connected to the functional core class and then every function. It retains syntactical data such as name, type of out value and eventually in-parameters.

Then pseudo-formal information ensures the safe level of the core. We made the choice of representing this information visually as it is shown on Figure 4b. This interface is the result of the compilation of the interfacing file and the functional core. We find it in the Builder module. This window is called *inspection window*. Each function is associated to a button, which may activate it interactively. Buttons are set on the left column, with function name inside. On the right column is indicated the type and the return value of the function. The type of the function determines whether it is a function which modify the functional core or a function that asks the functional core for its state. To characterize the identification of the functions, we classify them into two categories, **action** and **state** types.

**Action functions** are functions that modify the functional core state. For example, they can be called by the dialog control when the user wants to act on the final application. Action functions always have Boolean return value for feedback information onto the call.

To know the state of the functional core, the final application asks for **state functions**. Information state is returned by out values. The state functions may be evaluated by the final application at each action function call.

The Generator module determines the function type with the presence (action) or absence (state) of post-conditions in the pseudo-formal specification. For example, the function of Figure 3 is a state function because the pseudo-formal description does not contain any post-condition.

This distinction between action functions and state functions is not fundamental at this analysis level; it only helps in giving information to end-users about the functional core.

At this stage, an interactive application which permits to activate the functions that are usable according to the state of the functional core has been generated automatically. Functions without in-parameters may be activated in a way as simple as pushing a button. When there are in-parameters, they must be provided when calling the function. For that reason the generator builds for each function with parameters a dialog box that allows giving its parameters interactively (Figure 4f). With the pseudo-formal specification of each function, the generator checks for the validity of these conditions before calling the function.

At this step, we have a tool that allows testing the functional core and checking for correctness. This tool is completely automatically generated by the Generator tool. With help of semi-formal specifications we guarantee an important safety level. The part of the domain expert is now finished. The end-user may test the functional core with no fear of "crashing" it. More, he/she can now enhance the application by using the Builder module we study now.

### 3.2. Builder

The role of the "Builder" is to offer end-users a tool for developping easily their interactive application. This application uses the interfacing file generated by the Generator module and apply programming by demonstration and visual programming techniques. In this part we describe this module. Firstly we explain our choice concerning development tools of Builder application. Then, we deal with the tool itself by giving details about all the parts it is composed of. In a third part, we explain the way interface programming is done, with the example of the TicTacToe game. Finally we give the mechanisms which permit end-users to have some guarantees about safety.

**Tools**: The realization of our Builder tool needs an adapted toolbox with very strong interactive features. In fact, it must be able to create new interactive objects which hold direct manipulation, allowing at the same time direct activation of functional core actions. More precisely, the toolbox must authorize canceling any stage of compilation to test the final application. We chose the AMULET [6] toolbox. AMULET allows a dynamic administration of objects. In fact all is represented by objects with dynamic slots. Each slot has a name and may contain a typed value. It is possible to change dynamically the value or the type of any slot, and to add or cancel slots to any object. The second advantage of AMULET, for our purpose, is the administration of direct manipulation. It is possible to associate operations to each object, which correspond to different typical direct manipulation actions. These actions are represented by objects called interactors. AMULET has been realized to design easily an highly interactive interface for Win32, Unix and Macintosh systems. **GenBuild** has been developed first under Win32 system but was also tested on MacOS system and on Unix system.

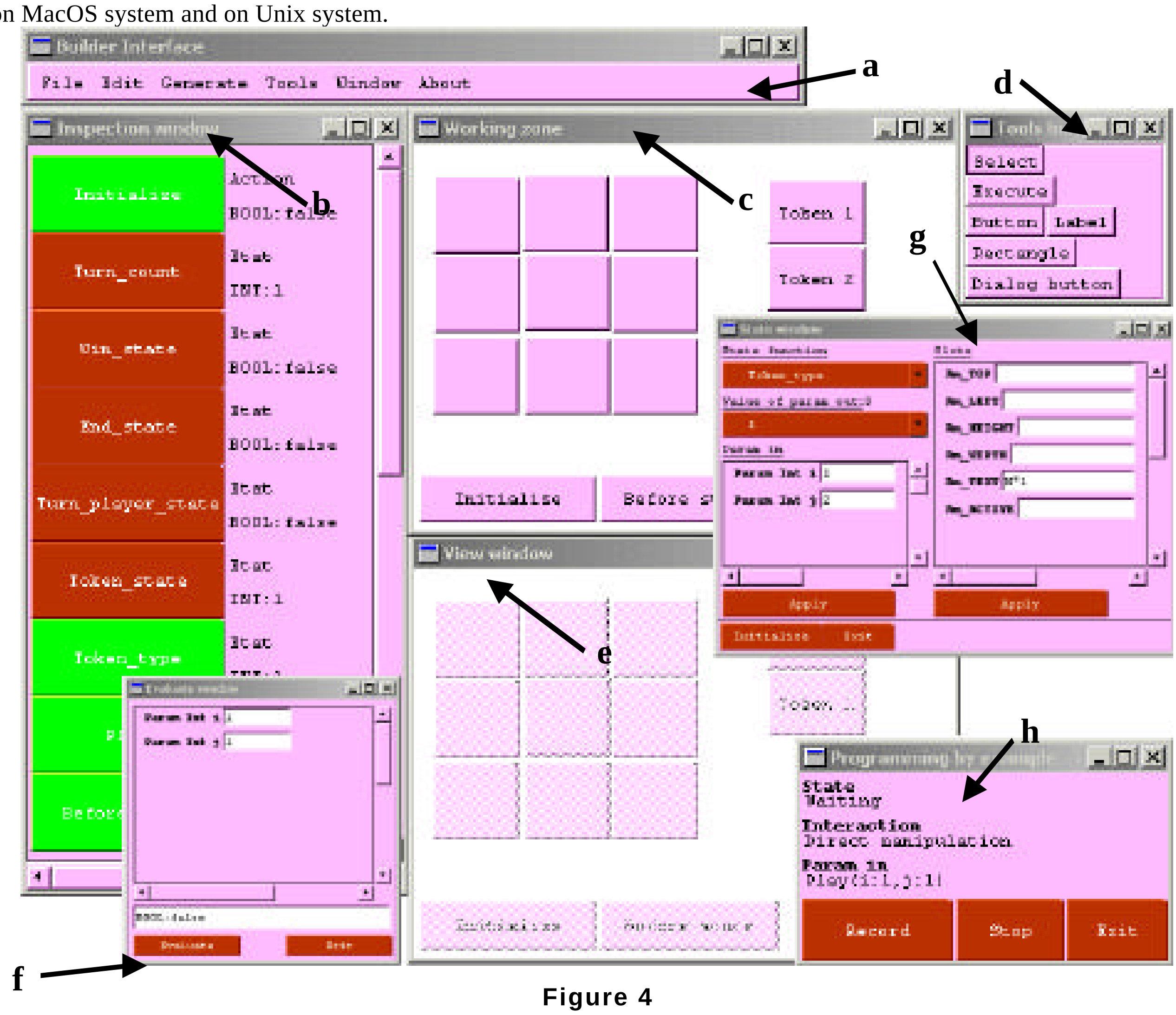

**Figure 4**

**(a) Tool bar – (b) Inspection window – (c) Working zone – (d) Tools box – (e) View window (f) Evaluate window – (g) State window – (h) Programming by example window**

**Builder module** : the Builder brings a tool to end-users that allows hime/her to easily generate his/her interactive application from a code generated by the Generator module. At this stage, the generation provides a model that permits testing the standard application, and then to avoid compilation (or interpretation) stage of traditional tools. This first approach simplifies yet the application design by end-user programming. End-users do not have to correct errors that would occur during compilation. They directly test the application. Then this method allows testing the application even if the program is not finished. Instead of leaving the automatically realized representation to modify it, we chose to furnish a design space that is independent from the inspection window. This solution permits to keep a visualization of the model state during the design. Another particularity of our tool is that the running context (during test phase) is not reinitialized or lost when the user switches to the design mode. So, it is easier for him/her to understand the state of his/her application.

The Builder is a graphic design tool, which proposes a friendly graphic working environment. It is composed of four windows, and one main toolbar (Figure 4a).

- The inspection window previously described (Generator module) remains active (Figure 4b);
- The working zone (Figure 4c) is the designer reserved space, for interface design. The end-user puts

objects from the tools palette on it. When this window is active, **GenBuild** is in design mode. The designer modifies the widgets appearance, size and position. He/She can program objects by displaying a popup menu with help of the right button of the mouse;

· The tools palette (Figure 4d) gives tools for manipulation and some AMULET widgets. The selection arrow takes the control of working zone objects. Then there are four AMULET objects which are widgets that the designer can put inside the working zone. *Button* is a classical object which modifies its appearance when the user clicks on it. *Label* is a text object allowing to show some text. *Rectangle* is a graphic object representing a rectangle and *dialog button* is a simple button followed by a confirmation dialog box;

· The visualization window (Figure 4e) represents the final application. When the design phase is on, all the objects which are created in the working zone are connected. For example, if the end-user moves an existing object into the working zone, the object is moved at the same time in the visualization window. Switching from the design stage to the test stage is done by clicking on the *execution button* on the tools palette. The working zone is then grayed, and it is not possible to modify the application anymore. During the execution stage the visualization window is connected to the inspection window. Any transformation will be transmitted to the inspection window. To come back to design mode, the user must push once again the *execution button*.

This is a first version of the Builder tool. It does not allow using all usual widgets, which would permit to use the Builder in order to develop any kind of application. For example it is not possible to create more than one application screen. The application may only be composed of one screen.

Let us now focus on the programming mechanisms, in order to see which techniques are given to end-users to develop safe interactive applications.

**Programming**: Application programming can be made into the working zone. The end-user puts widgets in the working zone and “programs” each of them. Two aspects may be programmed. The first is the interaction, and the second is the result that represents the state of the object after this interaction. For each of these two aspects, the end-user associates a functional core function by end-user programming. In fact, we used two programming techniques: the first one is visual programming, and the second one is programming by demonstration, with direct manipulation of the running example in the visualization window.

In all cases, effects of programming are the same. We used an important characteristic of AMULET, the dynamic administration of objects by adding or canceling slots. Programming consists in modifying a slot value, or in adding a new slot to an object.

· Interaction programming by visual programming. The Builder proposes to end-users many interactions on the objects of the interface: inputs from the mouse (left click, direct manipulation…) and inputs from the keyboard. Programming consists in choosing an interaction and associating it to an action function of the functional core by indicating the value of in-parameters. This development uses visual programming techniques and asks the end-user to repeat the same tasks. We established programming technique by demonstration in order to program more easily the interaction.

· Interaction programming by programming by demonstration. In order to explain the technique of programming by demonstration let us take an example on the TicTacToe game construction. The end-user wants to program the move of a piece on a grid element by direct manipulation. Because the behavior is the same for the nine squares, we have only to program one element with this technique. Its duplication may be done by simple copy/paste actions. At the beginning, the designer puts a rectangular object that represents a grid element, and he/she creates another rectangular object to assign a player piece. Then the end-user changes from **GenBuild** in programming by demonstration mode. This closes all windows but the design window. A new window appears (Figure 4h). This one controls the programming by demonstration process. It is a control panel, with buttons and textual information. A first button called *record* changes from a waiting state to record state. To cancel the programming by demonstration process when recording, the end-user pushes on the *stop button*. Then, a *close button* returns to classical programming. During the recording, **GenBuild** records intelligently all the interactions the designer wants to realize. The designer demonstrates to the system how to move a piece (Rectangular object) on the grid. The system detects the "collision" between objects, and warns the designer with a confirmation box. It asks him/her if the interaction found is the interaction the designer wanted to demonstrate. If the user confirms, the Builder asks for information about the function and about parameters to associate with this interaction. This is an intelligent help for interface programming. Completing the whole TicTacToe game consists in duplicating eight times this behavior.

· Action result programming. Finally the end-user must program the visualization of the actions result, which depends on the call result of state functions. This may be done with the dialog box (Figure 4g) which groups information to be modified. It is in fact the same procedure as the first method of interaction programming (associating a state function to AMULET object slots). The end-user chooses a state function among the whole functions of the functional core and for each return value, gives objects slots which are modified. For example the designer associates the type token function on a grid object. What the end-user wants to express is the type of the piece to be displayed in an element of the grid. He/She chooses firstly the type token function, the function parameters or for each return value he/she modifies the slots. By establishing this description we obtain the complete program.

**Safety aspects**: Let us now deal with the safety aspect of the application being developed with **GenBuild**. Security is omnipresent. It permits end-users to have a safe application development. First of all, we lead on a functional core defined in a semi-formal way. If it is well designed, we can be sure that the function calls are done without errors. The *Insistence* property may be checked on each function usage. Each function parameter is followed by an inspection post-condition which prevents from bad calls. It may also be tested whether each aspect of the semi-formal description results in modifying a slot.
The inspection window in the Builder environment gives a colored feedback about the function usage. So, the end-user can easily check if all functional core functions are used in the interface. This interface property is important because it permits to know if the interface the end-user has made is correct. This points deal with *Honesty* and *Observability* properties.

## 4. Related works

Many tools allow end-users to create interactive applications. For our purpose, we focus on tools that allow the design of stand-alone applications such as simulations or games. Some of them allow rather complex results. Firstly we can consider the Clickteamä products which all share the same fundamentals, and more precisely *The Games Factory* tool [7]. Designing application consists in putting objects on a *setting* and defining the behavior of each of them. In fact, each object is able to send out a fixed number of events. The designer must choose which events he/she wants to take into account and associate them to actions, which modify the interface state. This tool is laid out by a fixed set of events, objects and actions. When the user wants to program an event/action set that exceeds the tool capacities, he does not have any solution with end-user programming. *The Games Factory* only allows adding objects called "extension". These objects must be written in classical language.
*The Game Maker* [8] is similar to *The Games Factory* environment but the creation of interfaces is done through an object/instance approach. Each object is created using a frame. It has a fixed number of events and actions. The end-user chooses the actions he/she wants to associate to events and parameterizes each action. Actions use variables and functions inside the system and, similarly, the only way to improve programming is to create new functions in classical programming language. We can see clearly the interest to allow adding classical language functions in this kind of application. These two products own originally a fixed number of objects that permit to build applications. But if the conception has to be more customized, the lack of expressive power becomes a handicap. Moreover, the added code is not validated for safety. This solution might be compared with ours. The main difference can be found in the fact that no mechanism is given to the end-user in order to understand and to validate extensions usage. Our system of semi-formal descriptions was designed to solve this problem.
We can explore other tools which confine only to the interface programming by visual programming with no use of classical languages. *Stagecast Creator* [9] [10] [11] and even *Agentsheets* [12] are systems designed to create games or small simulations. The user creates worlds that contain objects with rules. A rule is defined by a starting context, and a final context. Programming consists in grouping several rules. At each time, the system checks the rules and apply the rule that matches the current context. These systems suffer from the problem of size: it is difficult to create large applications. The maintenance becomes quickly more and more complex. In fact the simple modification of the visual programming implies the modification of a whole part of the code. Moreover, the program visualization becomes harder and harder as the program size increases. Another example of this type of application is *Toontalk* [13]. The application development takes place as if the end-user played to make an application. The end-user visualizes the zone of development in 3D environment to the third person. *Toontalk* tools look like real construction tools (i.e. hammer) to simplify its comprehension. But this kind of application cannot elaborate high level interfaces and nor guarantee safety. If we rely to our work, we might consider the rules as a variant of our pre and post-conditions. The main difference is that rules are the only way to express programming in these tools, while it is only a part of our system. Adding events, either on simple objects or on couples of objects, enriches largely the end-user programming possibilities. Gamut is an environment of development which authorizes the complete design of an application by using techniques of programming by demonstration and programming by inference. The aim of Gamut [14] is to provide a tool that allows a complete development of interactive applications using programming by demonstration. Inference is largely used, with its strengths (usability) and its weaknesses (what can be done when inference does not find the right solution?). In fact, some options we offer to users rely to limited inference. It is the case, for example, when the system deduces the program from the direct manipulation of the example. Nevertheless, we do not address the same problem. Our main interest is safety. PBD systems generally prevent the user from making wrong actions, but is the action chain correct? This is the problem we focus on.

*PDGen* [15] is not devoted to game construction, but it may be related to our goals. It generates a complete stand-alone program, with a graphical interface to edit scientific data from files. From an analysis of $C^{++}$ code, and more especially on an analysis of classes definition, PDGen generates an interface that allows accessing data, navigating among them, and modifying them. The only aim of the program is this access to data, it is just possible to save modified data in a new file. PDGen may be viewed as a first step of our work. But, while our goal is to deal with any application domain, the semantic of PDGen applications is fixed. This system does not examine the member functions of classes, and does just know how to operate limited actions on data.

## 5. Conclusion and perspectives

We have presented the **GenBuild** tool, which permits an end-user to develop safe interactive applications. We are just in the first stage of our work. We have taken one limited example to experiment **GenBuild** feasibility. In the future we plan to extend this application to other case studies, for example in the process control domain.

**GenBuild** is composed of two distinct modules. The Generator is the first one. It is a specialized tool developped for a domain expert who sets out a safe functional core. By now, we are studying the eventuality of leading on a completely formal specification to realize our first generation. We are currently studying the use of a formal language (the B language) [16] to develop a proved functional core which would prevent the user from breaking application rules.

The Builder is the second module. It is a purely interactive tool that allows an end-user to develop a complete interactive application among an existing functional core. It allows the verification of some properties that are a first step towards the development of safe end-user programming. We have to improve this step, and also to consider task analysis as an help to usability checking. At this step, we must also pay attention to purely PBD systems such as Gamut [14] and examine how inference may be taken into account in our approach. It seems obvious that pre and post-conditions might help inference mechanisms in some cases.

## 6. References

[1] T. Lau, S. Wolfman, P. Domingos, and D. S. Weld, “Learning Repetitive Text-editing Procedures with SMARTedit,” in *Your Wish is My Command*, H. Lieberman, Ed., 2001, pp. 209-226.

[2] Y. Aït-Ameur, “Développements Contrôlés de Programmes par Modélisations et Vérifications de Propriétés,” in *LISI/ENSMA*. Poitiers: Université de Poitiers, 2000, pp. 146.

[3] C. Gram and G. Cockton, *Design Principles for Interactive Software*: Chapman & Hall, 1996.

[4] Pierra, *Les bases de la programmation et du Génie Logiciel*. Paris: Dunod informatique, 1991.

[5] J.-D. Fekete, “Les trois services du noyau sémantique indispensables à l'IHM,” presented at Journées Francophones sur l'Ingénierie de l'Interaction Homme-Machine (IHM'96), Grenoble, 1996.

[6] B. A. Myers, R. G. McDaniel, R. C. Miller, A. S. Ferrency, A. Faulring, B. D. Kyle, A. Mickish, A. Klimovitski, and P. Doane, “The Amulet Environment: New Models for Effective User Interface Software Development,” *IEEE Transactions on Software Engineering,*, vol. 23, pp. 347-365, 1997.

[7] Clickteam, (1996), "The Games Factory 1.06" , Clickteam & Europress Software, http://www.clickteam.com.

[8] M. Overmars, (2000), "Game Maker 3.2" , http://www.cs.uu.nl/~markov/kids/gmaker/.

[9] D. Canfield Smith, A. Cypher, and L. Tesler, “Novice Programming Comes of Age,” in *Your Wish is My Command*, H. Lieberman, Ed., 2001, pp. 7-20.

[10] A. Cypher and D. C. Smith, “KidSim: End User Programming of Simulations,” presented at Human Factors in Computing Systems (CHI'95), Denver, Colorado, 1995.

[11] D. Smith and A. Cypher, “KidSim : Child Constructible simulation,” presented at Imagina'95, Monte-Carlo, Février, 1995.

[12] A. Repenning and C. Perrone, “Programming by Analogous Examples,” in *Your Wish is My Command*, H. Lieberman, Ed., 2001, pp. 351-370.

[13] K. Kahn, “How Any Program Can Be Created by Working with Examples,” in *Your Wish is My Command*, H. Lieberman, Ed., 2001, pp. 21-44.

[14] R. G. McDaniel and B. A. Myers, “Getting More Out Of Programming by Demonstration,” presented at Human Factors in Computing Systems (CHI'99), Pittsburg, 1999.

[15] V. Engelson, D. Fritzson, and P. Fritzson, “Automatic generation of user interfaces from data structure specifications and object-oriented application models,” presented at ECOOP96, Linz Austria, 1996.

[16] Steria Méditerranée, (1997), "Atelier B 3.5" .